\documentclass[11pt]{article}
\usepackage{amssymb,amsmath}
\usepackage{latexsym}
\usepackage{color}
\usepackage{epsfig}
\newcommand\p\partial

\newcommand{\bea}{\begin{eqnarray}}
\newcommand{\eea}{\end{eqnarray}}
\newcommand{\be}{\begin{equation}}
\newcommand{\ee}{\end{equation}}

\begin{document}



\begin{flushright}
NSF-KITP-10-088
\end{flushright}
\vskip 1in

\begin{center}
{\Large \bf 
Non-Einstein geometries in Chiral Gravity}
 \vspace*{0.5cm}\\
 {Geoffrey Comp\`ere$^{\flat}$, Sophie de Buyl$^{\flat}$ and St\'ephane Detournay$^{\natural}$
 }
\end{center}
 \vspace*{0.1cm}
 
 \begin{center}
 $^{\flat}${\it Department of Physics, 
 University of California, Santa Barbara, \\ 
 Santa Barbara, CA 93106, USA\\
  {\tt gcompere@physics.ucsb.edu}$\quad$ 
 {\tt sdebuyl@physics.ucsb.edu} 
 } 
 \vspace*{0.5cm}\\
 $^{\natural}${\it Kavli Institute for Theoretical Physics, \\ 
 University of California, Santa Barbara, CA 93106, USA\\
 {\tt detourn@kitp.ucsb.edu}
 } 
 \vspace{0.15cm}
 
 \end{center}

 \vspace{0.15cm}

 \begin{abstract}

We analyze the asymptotic solutions of Chiral Gravity (Topologically Massive Gravity at $\mu \ell =1$ with Brown-Henneaux boundary conditions) focusing on non-Einstein metrics. A class of such solutions admits curvature singularities in the interior which are reflected as singularities or infinite bulk energy of the corresponding linear solutions. A non-linear solution 
is found exactly. The back-reaction induces a repulsion of geodesics and a shielding of the 
singularity by an event horizon but also introduces closed timelike curves.

%


 \vspace{12pt}
 Pacs:  04.20.-q,04.60.-m,04.70.-s,11.30.-j
 \vspace{12pt}
 \end{abstract}

Chiral gravity has recently been proposed as a consistent, non-trivial and stable pure gravity theory in three dimensions containing black holes \cite{Li:2008dq}, which would make it an extremely valuable toy model to study various aspects of quantum gravity.
This theory is a special case of Topologically Massive Gravity (TMG) \cite{Deser:1981wh,Deser:1982vy,Carlip:2008eq} at a particular point in parameter space and with phase space defined by asymptotically $AdS_3$ Brown-Henneaux boundary conditions \cite{Brown:1986nw}. The chirality of the theory at the classical level manifests itself through its asymptotic symmetry group consisting in a single copy of a Virasoro algebra acting on the phase space of the theory \cite{Strominger:2008dp,Henneaux:2009pw}. It has furthermore been conjectured \cite{Li:2008dq} that a positive 
energy theorem holds for the theory. 

At the linearized level, examples were proposed in the literature disproving positivity \cite{Carlip:2008jk,Giribet:2008bw,Park:2008yy,Carlip:2008eq,Carlip:2008qh}. However, it was subsequently shown that they all suffer from linearization instabilities \cite{Maloney:2009ck}. Defining a theory with consistent massive graviton excitations
at the chiral point instead requires to relax the Brown--Henneaux boundary conditions allowing for a logarithmic term,
as described in \cite{Grumiller:2008pr,Grumiller:2008qz,Grumiller:2008es}. This leads to a theory that is dual to a logarithmic CFT \cite{Grumiller:2008pr,Grumiller:2009mw}. At the non-perturbative level, it was proven that all stationary, axially symmetric solutions of chiral gravity are the familiar  BTZ black holes \cite{Banados:1992gq} from pure Einstein gravity \cite{Maloney:2009ck}. Whether all solutions of chiral gravity are Einstein metrics had been left as an open question.

At the quantum level, it was argued that the partition function computed by summing over real saddle points including all perturbative corrections is the chiral half of the extremal partition function proposed by Witten as the dual of pure 3d gravity \cite{Witten:2007kt, Maloney:2007ud, Maloney:2009ck}. The latter real saddle points turned out to satisfy euclidean Einstein's equations.

The purpose of this note is to show that there exists solutions to chiral gravity which do not satisfy Einstein's equations. These solutions are non-stationary or non-axisymmetric in accordance with the Birkoff theorem proven in \cite{Maloney:2009ck}. Our analysis will be based on the recent work \cite{Skenderis:2009nt} which determined the most general asymptotic solution of TMG at the chiral point in Fefferman-Graham form. As we will see in Sec. 1, the equations of motion of TMG are less restrictive than in pure gravity, and allow for more general asymptotic solutions. The latter are in general expressed as an infinite asymptotic series at the boundary. We should however worry about the regularity of the solution in the interior geometry in order to know if we should regard them as physical or not. We address that issue from two different perspectives. In Sec. 2, we analyze a large class of linear perturbations around AdS and the BTZ black holes absent in pure Einstein gravity but obeying Brown-Henneaux boundary conditions. We find that those perturbations are singular and have infinite bulk energy. We analyze in Sec. 3 the effects of back-reaction on one particular perturbation. The non-linear solution admits a curvature singularity which however coincides with a past event horizon. The singularity is therefore shielded. However, naked closed timelike curves are also present in the interior geometry and allow causality violation. We will conclude with a discussion on the implications of these solutions on classical and quantum chiral gravity.

\section{Asymptotic solutions of Chiral Gravity}

Asymptotically anti-de Sitter space times can be defined as conformally compact manifolds whose boundary metric is conformal to a cylinder. Such spacetimes admit an asymptotic Fefferman-Graham expansion whose precise form depend on the equations of motion of the theory considered. In Topologically Massive Gravity (TMG) at $\mu \ell =1$, such an expansion was determined in \cite{Skenderis:2009nt} to be\footnote{We use the conventions of \cite{Li:2008dq} and $\epsilon^{r t\phi}= +1$.}
\begin{equation}
ds^2 = \frac{dr^2}{r^2} + (-r^2 dt^2 + r^2 d\phi^2) + \left(\log{r} b_{(2)ij}+ g_{(2)ij} + o(r^0) \right) dx^i dx^j
\end{equation}
where $b_{(2)ij}$ and $g_{(2)ij}$, $i=t,\phi$ are further constrained by the equations of motion. Chiral gravity can be defined as the subset of this theory where the logarithmic branch is turned off, 
\be
b_{(2)ij} \equiv 0,
\ee
which can be done consistently with the equations of motion. This condition is equivalent to setting the left set of Virasoro charges to zero \cite{Maloney:2009ck} and reduces the phase space to metrics obeying the Brown-Henneaux boundary conditions \cite{Brown:1986nw}. At second order in the Fefferman-Graham expansion, the equations of motion imply that $g_{(2)}$ is traceless and that a \emph{chiral projection of} its divergence is fixed. In Einstein gravity, the whole divergence of $g_{(2)}$ is fixed \cite{deHaro:2000xn}, but in TMG at the chiral point, one condition is relaxed and only a chiral half of the divergence is fixed by the equations of motion \cite{Solodukhin:2005ah}. The general non-linear solution at second order is then
\begin{equation}
g_{(2)ij} =  \left( \begin{array}{cc} 1 & 1 \\ 1 & 1 \end{array} \right)_{ij} F(t,\phi) + 
\left( \begin{array}{cc} 1 & -1 \\ -1 & 1 \end{array} \right)_{ij} \bar L(t-\phi).\label{g2}
\end{equation}
It is easy to check that the modes of the function $\bar L(t-\phi)$ are in one-to-one correspondence with the eigenvalues of the right-moving Virasoro charges while the function $F(t,\phi)$ does not appear in the asymptotically conserved charges, see the definitions of charges e.g. in \cite{Henneaux:2009pw}. Also, the holographically renormalized stress-tensor \cite{Skenderis:2009nt,Ertl:2009ch} 
\be
T_{\phi\phi} = T_{tt} = {1 \over 2 G} \bar L, \qquad T_{t\phi} = -{1 \over 2 G}Ê\bar L
\ee
does not depend on $F$. We observe that the Cotton tensor is zero at second order in $r$ if and only if $F(t,\phi)$ is a left-moving function $L(t+\phi)$ corresponding to solutions of Einstein gravity.
The novelty in Chiral gravity with respect to Einstein gravity comes therefore from the part of the function $F(t,\phi)$ which is not of the form $L(t+\phi)$. Such a function cannot be removed by a diffeomorphism. As a reminder, the two functions $\bar L(t-\phi)$ and $L(t+\phi)$ except their zero modes can be removed upon acting with a globally defined large diffeomorphisms generated by the two Virasoro algebras. The zero modes of $L(t+\phi)$ and $\bar L(t+\phi)$ are equal to the two parameters of the BTZ, see e.g. eq (1) of \cite{Banados:1998gg}.

The asymptotic solutions with $F(t,\phi)\neq 0$ can be solved at higher orders with the following expansion
\begin{equation}
ds^2 = \frac{dr^2}{r^2} + (-r^2 dt^2 + r^2 d\phi^2) + \left( g_{(2)ij} +\frac{1}{r^2} g_{(4)ij} +\frac{1}{r^4} g_{(6)ij} + \dots \right) dx^i dx^j \label{asymptsol}
\end{equation}
Using the differential operator $D_a \equiv \p_t + a\p_\phi$, one finds 
\begin{eqnarray}
g_{(4)ij} &=&  
\frac{1}{16}\left( \begin{array}{cc} -D_{-3}& D_1 \\ D_1 & 3D_{-1/3} \end{array} \right)_{ij} D_{-1}F(t,\phi)+\left( \begin{array}{cc} -1 & 0 \\ 0 & 1 \end{array} \right)_{ij} F(t,\phi)\, \bar L(t-\phi),\\
g_{(6)ij} &=&  \frac{1}{576}\left( \begin{array}{cc} -D_{-1+\sqrt{6}}D_{-1-\sqrt{6}}&D_{2+\sqrt{3}}D_{2-\sqrt{3}}  \\ D_{2+\sqrt{3}}D_{2-\sqrt{3}}  & 5D_{(1+\sqrt{6})/5}D_{(1-\sqrt{6})/5}  \end{array} \right)_{ij}D_{-1}^2 F\nonumber\\
&+&\frac{1}{8} \left( \begin{array}{cc} 1 & 1\\ 1 & 1 \end{array} \right)_{ij} (F D_{-1}-\frac{1}{12}D_{-1}F)D_{-1}F 
+\frac{1}{72}\left( \begin{array}{cc} 1 & -1 \\ -1 & 1 \end{array} \right)_{ij} \bar L^\prime  D_{-1}F,\nonumber\\
&+&\frac{1}{72} \bar L \left( \begin{array}{cc} - D_{17}& -8D_{-1} \\ -8D_{-1} & 17 D_{1/17} \end{array} \right)_{ij} D_{-1}F\,\nonumber \\
g_{(8)ij} &=& \dots 
\end{eqnarray}
In general, the Fefferman-Graham expansion does not terminate. It terminates at order 4 ($g_{(6)ij}$ and higher are zero) for all solutions of Einstein gravity \cite{Skenderis:1999nb} where $D_{-1}F = 0$. 

A solution of chiral gravity has to obey regularity conditions in the interior. If curvature singularities are present, they should be shielded by an horizon. We now turn our attention to that issue by studying first the linearized theory. 

\section{Linear perturbations}

Studying the physical content of the entire class of metrics \eqref{asymptsol} is a tedious task. In order to grasp some insight we will study linear perturbations around anti-de Sitter space and around the BTZ black hole and investigate whether these linear solutions are regular in the interior geometry and have finite energy.

Anti-de Sitter space and the BTZ have the form \eqref{asymptsol} where the only non-vanishing Fefferman-Graham coefficients are
\begin{equation}
g_{(2)ij} = 
\left( \begin{array}{cc} m+\alpha& \alpha-m \\ \alpha - m & m+\alpha   \end{array} \right)_{ij} \quad, \quad
g_{(4)ij} = 
\left( \begin{array}{cc} -m\alpha & 0 \\ 0 & m\alpha \end{array} \right)_{ij} \, .\label{BTZ}
\end{equation}
Anti-de Sitter space is given by $m = \alpha = -1/4$ while the BTZ black holes have $m,\alpha \geq 0$. The Fefferman-Graham coordinate $r$ extends from infinity to $r = 1/2$ for AdS where the spacetime ends. For the black holes, the coordinate $r$ reaches the horizon at $r_+= (m \alpha)^{1/4}$. These bounds are more easily seen by using the standard BTZ coordinate $R = (r^2 + (m+\alpha)+m\alpha r^{-2})^{1/2}$ such that $g_{\phi \phi}  = R^2 d\phi^2$. For the sake of completeness, the Virasoro zero modes are given by
\be
L_0 = 0,\qquad \bar L_0 = \frac{m}{2G} . 
\ee
As a starter, one could try to find if highest-weight solutions exist in the Fefferman-Graham gauge as they do in traceless-transverse gauge \cite{Li:2008dq}. In Fefferman-Graham coordinates, the $SL(2,\mathbb R)_L$ generators have the form
\begin{eqnarray}
L_0 &=& \frac{i}{2}(\p_t +\p_\phi)\\
L_{-1} &=& \frac{i}{2} e^{-i (t+\phi)} \left( \frac{4r^2-1}{4r^2+1}\p_t + \frac{4 r^2+1}{4r^2-1} \p_\phi + i r\p_r \right)\\
L_{1} &=& \frac{i}{2} e^{i (t+\phi)} \left( \frac{4r^2-1}{4r^2+1}\p_t + \frac{4 r^2+1}{4r^2-1} \p_\phi - i r\p_r \right).
\end{eqnarray}
and obey $[L_1,L_{-1}] = 2L_0$, $[L_{\pm 1},L_0]= \mp L_{\pm 1}$. The $SL(2,\mathbb R)_R$ generators $\bar L_n$ are obtained by inverting $\phi \rightarrow -\phi$ in the above. Starting from the anzatz 
\begin{eqnarray}
h_{\mu\nu}dx^\mu dx^\nu = e^{-i (h+\bar h) t + i (h - \bar h) \phi} (g_1(r) d\phi^2 + 2 g_2(r) dt d\phi + g_3(r) d\phi^2)
\end{eqnarray}
and imposing that the perturbation is invariant under $L_{-1}$ and $\bar L_{-1}$, one finds that $h = \bar h $ and $g_1 \sim g_2 \sim g_3 \sim (r^{-1}+4r^2)^{2h}$. The equations of motion then fix $h = 1$ and $g_1 = g_2 = 0$. The linearized solution do not obey the Brown-Henneaux boundary conditions and is therefore not tangent to the phase space of chiral gravity. There are therefore no highest weight state perturbations in the Fefferman-Graham gauge. We therefore turn our attention to perturbations which do not fall in highest weight representations.

Let us consider the following anzatz 
\be
F(t,\phi) = \alpha + \epsilon t ,\qquad \bar L(t-\phi) = m.
\ee
Keeping only terms linear in $\epsilon$ in the equations of motion, we obtain the following linear solution 
\begin{equation}
h_{\mu\nu}dx^\mu dx^\nu =  t (dt + d\phi)^2 + \frac{m t}{r^2} (-dt^2 + d\phi^2) \label{linmode}
\end{equation}
which admits a finite Fefferman-Graham series expansion. The solution is linearly divergent in time and might potentially represent an instability of the background if it is a normalizable and bounded perturbation, see e.g. discussions in \cite{Gibbons:2002pq}. The perturbation is manifestly not bounded when t goes to infinity. However, since the $(r,t)$ coordinate system breaks down at the black hole horizon, one has to re-express the perturbation in Kruskal coordinates $(U,V,\phi)$ (see \cite{Banados:1992gq}) to be sure this is not a coordinate artifact. We checked that the perturbation indeed blows up at the black hole horizon for the cases $m = \alpha$ and $m,\alpha >0$, and therefore does not represent a genuine instability. This divergence suggests that any non-linear completion of the perturbation \eqref{linmode} of the form $g = \bar g + h + h_{(2)}+\dots $ admits curvature singularities. Another argument for the irregularity of the perturbation goes as follows. The equations of motion $E_{\mu\nu}\equiv \sqrt{-g}(G_{\mu\nu}-g_{\mu\nu}+C_{\mu\nu})$ admit the perturbative expansion $E_{\mu\nu}[g]=E_{\mu\nu}[\bar g]+E^{(1)}_{\mu\nu}[h ;\bar g]+E^{(2)}_{\mu\nu}[h,h ; \bar g]+ E^{(1)}_{\mu\nu}[h_{(2)};\bar g ]+\dots$. The energy of the mode \eqref{linmode} can be computed using a bulk integration as
\begin{equation}\label{BulkE}
Q_{bulk}(\p_t) = \frac{1}{16 \pi G} \int_\Sigma *((\p_t)^\mu E_{\mu \nu}^{(2)} (h,h ; \bar g) dx^\nu),
\end{equation}
where $\Sigma$ is a constant time slice. Now, it turns out that this energy always diverges due to the inner bound of the integration either at the origin of AdS $R_0 =0$ or at the horizon of BTZ $R_0 = \sqrt{m}+\sqrt{\alpha}$. Indeed, we find
\begin{eqnarray}
Q_{bulk}(\partial_t)& \sim& \int_{R_0}^\infty \frac{dR}{(R-R_0)^{d+1}} \left( 1 + O(R^{1}) \right) \sim \infty \qquad \text{for $d >0$}
\end{eqnarray}
where the degree of divergence $d$ (defined with respect to the coordinate $R$) is always positive. Its explicit value turns out to depend on the background metric in an non-trivial manner, see Table~\ref{table1}. 

\begin{table}[tbh]
\begin{center}
\begin{tabular}{c|c|c|c|}
Background geometry & $\alpha$ & $m$ & degree of divergence \\ \hline 
AdS &$-1/4$ & $-1/4$ & 1 \\
non-extremal BTZ & $>0$ & $>0$& $3/2$\\
Extremal BTZ with $\bar L_0 = 0$ & $>0$ & 0& 2\\
Vacuum BTZ & 0 & 0& 4\\
Extremal BTZ with $\bar L_0 > 0$ & 0 & $>0$ & 5
\end{tabular}
\end{center}\caption{Degree of divergence of the energy of the linearized solution linear in time in the interior of the background geometry at $R = R_0$.}\label{table1}
\end{table}

One can generalize the above analysis to the anzatz 
\be
F(t,\phi) = \epsilon\; t^n ,\qquad \bar L(t-\phi) = m \; \qquad n \geq 2
\ee
with a polynomial function of time of degree 2 or higher. We choose to analyze the solutions around the background $\alpha = 0$ for simplicity. It turns out that the linear solution has a finite Fefferman-Graham expansion which stops at $r^{-2 \lfloor \frac{n}{2} \rfloor}$ around the vacuum BTZ $(m=0)$ and at $r^{-4 \lfloor \frac{n}{2} \rfloor  -2 }$ around the extremal BTZ with $\bar L_0 >0 $ $(m > 0)$. The linear solution takes the form
\begin{eqnarray}
h_{\mu\nu}dx^\mu dx^\nu =  \sum_{i=0}^{\lfloor n/2 \rfloor} \frac{n!}{2^{2i}((i+1)!)^2 (n-2i)!}\frac{t^{n-2i}}{r^{2i}} \left( \sum_{j=0}^{i+1} ds^2_{(i,j)} \frac{m^j}{r^{2j}}\right)\label{linsoln}
\end{eqnarray}
where 
\begin{eqnarray}
&&ds^2_{(0,0)} = (dt+d\phi)^2, \qquad ds^2_{(0,1)} = -dt^2 + d\phi^2,\nonumber\\
&&ds^2_{(1,0)} = -dt^2 + 2dt d\phi+d\phi^2,\quad   ds^2_{(1,1)} = \frac{2}{9} (-dt + 17d\phi)(-dt +d\phi),\quad ds^2_{(1,2)}= (dt-d\phi)^2,\nonumber\\
&&ds^2_{(i,0)} = -dt^2 + 2dt d\phi+(2i+1)d\phi^2,\quad \dots\;,\quad ds^2_{(i,i+1)}= \frac{1}{Cat(i)}(dt-d\phi)^2 \qquad \forall i \geq 1
\end{eqnarray}
where $Cat(i)= \frac{2i!}{i!^2 (i+1)}$ are the Catalan numbers. The general expression $ds^2_{(i,j)}$, $0 <j \leq i$ can be expressed in terms of intricate combinatorial factors that we do not need to write down here for our arguments. We checked that \eqref{linsoln} is a linear solution around the background \eqref{BTZ} with $\alpha =0$ and $m$ arbitrary for all $n \leq 15$. 

The linear perturburation is always singular at $r=0$ which is at the horizon of the massive BTZ black hole $(m >0)$ or at the origin of the vacuum BTZ $(m=0)$. We therefore expect that any non-linear completion of this perturbation will be singular. Further evidence comes from the bulk energy 
\begin{eqnarray}
Q_{bulk}(\partial_t)& \sim& \int_{\sqrt{m}}^\infty \frac{dR}{(R-\sqrt{m})^{3+2n}} \left( 1 + O(R^{1}) \right) \sim \infty \qquad \text{for all $n \geq 2$, $m \geq 0$}
\end{eqnarray}
which always diverges with a degree of divergence $d = 2 n +2 \geq 6$ for $n \geq 2$ superior to the ones of the perturbation linear in time. In that sense, the polynomial solutions are more and more singular with increasing $n$. The divergence in the energy here does not signal a breakdown of the Brown-Henneaux boundary conditions. Rather, it signals the existence of a singularity in the interior geometry.

We have explored a restricted set of non-Einstein linearized perturbations, namely a subset of which have a finite Fefferman-Graham expansion. In all cases analyzed, the perturbations are singular and have infinite bulk energy. Moreover, the divergence in the energy increases qualitatively for polynomial modes in time of higher order. We don't expect that adding an angular dependence in $F(t,\phi)$ will resolve these singularities. In order to prove or disprove the conjecture that the only regular solutions of chiral gravity are Einstein metrics, one would have to analyze the entire class of linear solutions with $F(t,\phi)$ arbitrary. It seems to be a hard problem that we leave unsolved. 

A caveat of the above analysis is that even if solutions of chiral gravity are irregular, the singularities might be hidden by an horizon introduced by back-reaction effects. In order to study that possibility we now turn our attention to a particular case where the back-reacted solution can be found. 

\section{Back-reaction effects: repulsion, horizon shielding but closed timelike curves}

One can ask if the presence of a singularity that we infered in the last section from the linear analysis could not be hidden by an horizon in the non-linear solution when all higher order perturbative corrections are taken into account. We will develop the answer to that question for the most simple example of the linear time perturbation \eqref{linmode} around the extremal BTZ background with $\bar L_0 = 0$ $(m=0)$ in what follows. First, we note that in this simple case the perturbative expansion stops at second order. The non-linear solution $g = \bar g + h + h_{(2)}$ is given by
\begin{equation}
ds^2 = \frac{dr^2}{r^2} + (-r^2 dt^2 + r^2 d\phi^2) + \left( \alpha + t \gamma - \frac{\gamma^2}{96 r^4} \right) (dt + d\phi)^2 .\label{eq:t}
\end{equation}
This solution falls into the class \eqref{asymptsol} with $g_{(6)} \neq 0$ and no higher Fefferman-Graham coefficients. We set the parameter $\alpha$ to zero by a shift of time. The conserved charges of this solutions are simply
\be
L_0 = 0,\qquad \bar L_0 = 0.
\ee
At $r=0$, the system of coordinates breaks down but two physical effects can be seen: the norm of the Killing vector field ${\partial \over \partial \phi}$ is infinite and the Ricci tensor contracted two times with ${\partial \over \partial \phi}$ is infinite as well. If $r=0$ can be reached by geodesics in finite affine time, it represents a physical singularity. Indeed, a vector field on a regular manifold cannot have infinite norm except asymptotically close to its open boundaries. However, this singularity might be hidden if geodesics cannot reach $r=0$ in finite coordinate time $t$. Another source of concern is the presence of closed timelike curves in the spacetime region beyond the velocity of light surface (VLS) where the norm of $\p_\phi$ becomes null,
\be
r^2+ t \gamma - \frac{\gamma^2}{96 r^4} = 0 \qquad (VLS).\label{vls}
\ee
One could ask whether geodesics can probe or not that region and reach the AdS boundary and whether the coordinate time always increases on timelike and null geodesics. 

In order to settle those questions, we have to solve the geodesics equations. Since the spacetime is time-dependent these equations do not have a simple form and have to be solved numerically. Nevertheless, before doing so, one can get some insights in the role of the back-reaction term $h_{(2)}$ by considering null geodesics around the spacetime
\begin{equation}
ds^2 = \frac{dr^2}{r^2} + (-r^2 dt^2 + r^2 d\phi^2) - \frac{\hat\gamma^2}{r^4}  (dt + d\phi)^2 .\label{eq:ts}
\end{equation}
which is a valid approximation to the near $r=0$ region of \eqref{eq:t} with $\hat \gamma = \gamma/(4\sqrt{6})\neq 0$ when $t$ is finite. Denoting by $E = \dot x^\mu g_{\mu t}$ and $J = \dot x^\mu g_{\mu \phi}$ the conserved energy and angular momentum along the null geodesic $x^\mu(\tau)$, we find that
\be
\dot{r}^2 = (E^2 - J^2) - \frac{\hat \gamma^2}{r^6}(E-J)^2\label{firstorder}
\ee
where the dot denotes differentiation with respect to affine time $\tau$. Since $\dot{r}^2 \geq 0$, geodesics admit a solution only for $E^2 > J^2$ or $E = J$. For $E = J$, the null rays lie on circular orbits at fixed radius while for $E^2 > J^2$ null rays reach a minimum radius and are repulsed to infinite radius as can be seen from the second order differential equation for $r$
\be
\ddot{r} = \frac{3\hat \gamma^2 (E-J)^2}{r^7}.
\ee
The back-reaction term can therefore be thought of as inducing a repulsive barrier close to $r=0$.

Let us now go back to the solution of chiral gravity \eqref{eq:t}. The repulsion of geodesics will take place as long as the time coordinate $t(\tau)$ does not diverge when $r(\tau)$ approaches zero. When $t(\tau)$ diverges, the approximation \eqref{eq:ts} is not valid and one has to solve the complete geodesic equations. However, in the later case, the singularity will be in the future or past event horizon of an observer at infinity. This effect can be probed more directly by noting that the two equations $J = \dot x^\mu g_{\mu \phi}$ and $g_{\mu\nu}\dot{x}^\mu \dot{x}^\nu =0$ imply as a consequence of $\dot t^2 \geq 0 $ that 
\be
\gamma t \geq \frac{\gamma^2}{96 r^4} - r^2 \left( \frac{J^2}{\dot{r}^2}+1 \right).\label{constr}
\ee
When $\gamma = 0$, there is no constraint. Also, at infinity there is no constraint, consistently with the behavior of geodesics in AdS. The relation now implies that all geodesics crossing $r=0$ will necessarily reach $t = +\infty$ when $\gamma >0$ and $t = -\infty$ when $\gamma <0$. This points to the existence of a past event horizon and a future event horizon for $\gamma >0$ and $\gamma < 0$, respectively. Once the choice of time flow in the boundary AdS metric is fixed, the solution with $\gamma$ positive or negative are physically inequivalent. In the former case the singularity lies in the future and cannot influence the past, while in the latter case the singularity lies in the past and influences the asymptotic observer. The latter spacetime is pathological since the singularity $r=0$ is naked. We will therefore only consider $\gamma >0$ in what follows. The curvature singularity is then safely hidden beyond a past event horizon. In that sense, the back-reaction resolves the singularity found in the linear theory by shielding it with an horizon.

\begin{figure}[hbt]
\begin{center}
\resizebox{0.5\textwidth}{!}{\mbox{\includegraphics{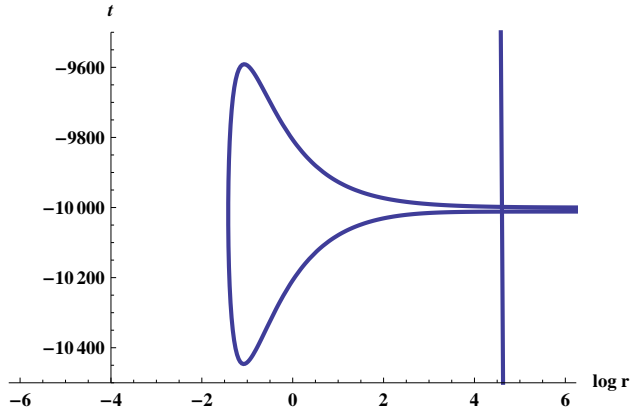}}}
\caption{Initially future-directed null geodesic with $J = -200$, $t_0 = -10000$. The geodesic comes from AdS on the upper part of the right-hand side, enters the CTC region after crossing the velocity of light surface plotted by the vertical line, and goes back to the AdS region at a lower coordinate time $t$ thereby violating causality.}\label{fig:CTC}
\end{center}
\end{figure}

Let us now study in more details the closed timelike curves. Let us pick up a point $p$ lying beyond the velocity of light surface \eqref{vls} but outside the singularity $r_p > 0$. The norm of $\p_\phi$ at that point is given by $-c^2$ with $c>0$. Choosing for convenience the proper time of the geodesic such that $\dot{r}(p) = \pm 1$, the constraint \eqref{constr} then enforces that 
\be
c \leq r_p |J|.
\ee
We see that geodesics without angular momentum cannot probe the region with closed timelike curves. The constraint \eqref{constr} is then exactly the condition that geodesics cannot penetrate inside the velocity of light surface. However, nothing prevents to probe regions with arbitrary negative norm $\p_\phi$ with geodesics with a high enough angular momentum. An example is presented using the numerics in Fig \ref{fig:CTC}. The closed timelike curves are therefore naked and we conclude that the spacetime \eqref{eq:t} is not very physical.

Let us present the numerical results. Using a rescaling of the coordinates and a change of periodicity of $\phi$ (that will not affect any of the results here), one can set $\gamma = +1$. We will integrate the geodesic starting from the AdS boundary $r = \infty$ at $\tau =0$. We can rescale the affine time $\tau$ such that $\dot{r}(0) = -1$ and we fix $\phi(0)= 0$. The initial data of the geodesic then only reduces to the angular momentum $J$ and to the initial time $t(0) = t_0$ both taking values in $\mathbb R$, and, the sign of the time flow $\dot t(\infty)$.

\begin{figure}[thb]
\begin{minipage}{0.45\textwidth}
\begin{flushleft}
\resizebox{\textwidth}{!}{\mbox{\includegraphics{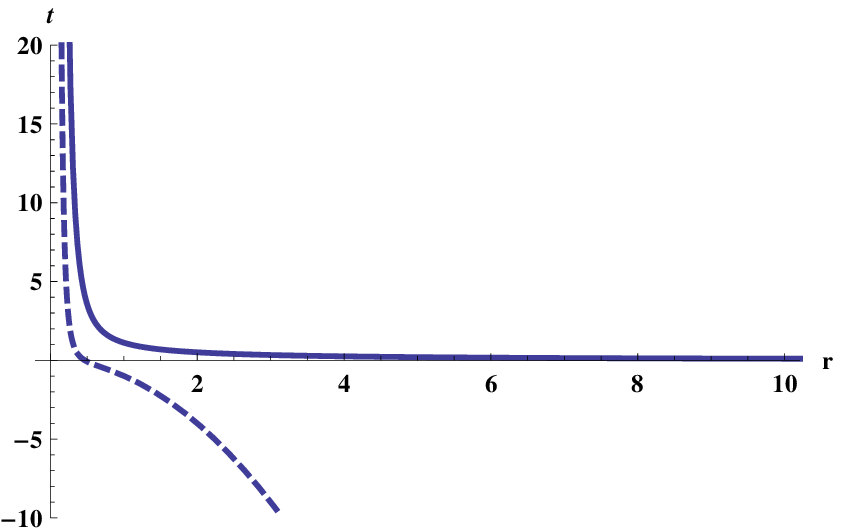}}}
\caption{Future-directed null geodesic with $J = 0$, $t_0 = 0$. The geodesic comes from AdS on the right-hand side and goes towards the singular horizon $r=0$, $t =\infty$. The region to the left of the dashed curve (VLS) is the region with CTCs forbidden by \eqref{constr}.}
\end{flushleft}
\end{minipage}
\begin{minipage}{0.10\textwidth}\mbox{}$\qquad$
\end{minipage}
\begin{minipage}{0.45\textwidth}
\begin{flushright}
\resizebox{\textwidth}{!}{\mbox{\includegraphics{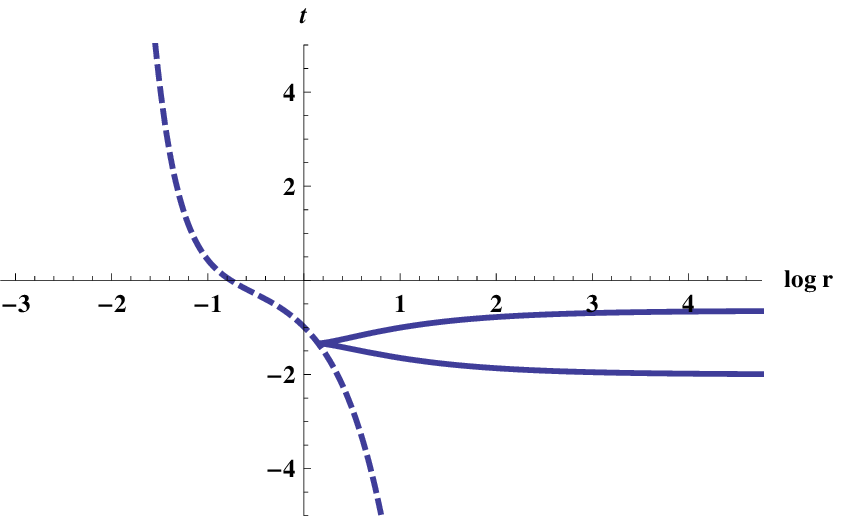}}}
\caption{Future-directed null geodesic with $J = 0$, $t_0 = -2$. The geodesic comes from AdS on the right-hand side and goes back to AdS after bouncing at positive radius. The region to the left of the dashed curve (VLS) is the region with CTCs forbidden by \eqref{constr}.}
\end{flushright}
\end{minipage}
\end{figure}

At zero angular momentum, the geodesics fall into two distinct classes depending if $t_0$ is bigger or smaller than the numerically found critical value
\be
t_* = -1.4165 \pm 0.0001
\ee
When $t_0 > t_*$, future-going geodesics reach $r = 0$ in finite affine parameter. The time coordinate $t$ diverges there and $\p_\phi$ has positive norm. The geodesic reaches the singularity. When $t_0 < t_*$ future-going geodesics reach a minimal positive radius and bounce back to the AdS region. The geodesics can be extended to infinite affine parameter and the coordinate $t$ reaches a finite value, as it should in AdS. 

When the angular momentum is turned on one can find numerically a critical time $t_*(J) < 0$ such that geodesics either reach the horizon $r=0$ at $t=\infty$ or bounce back. The function $t_*(J)$ is not symmetric upon flipping the sign of $J$ as a consequence of the metric being parity violating.
Most importantly, there exists null geodesics probing the CTC region and going back to the AdS region at an earlier time, thereby violating causality. An example of such a geodesic is shown in Fig \ref{fig:CTC} as previously announced.

\section{Discussion}

The existence of non-Einstein solutions to chiral gravity raises a certain number of questions, both on the classical and quantum consistency of chiral gravity. We argued that curvature singularities could be present for all solutions. Such singularities might be hidden by an horizon, as we showed on one example, but other pathologies, like closed timelike curves, possibly render these solutions unphysical. We leave as an open classical problem the interested reader to show the (non-)existence of regular non-Einstein solutions to chiral gravity. It is interesting to observe that the structure underlying these solutions is specific to the chiral point $\mu l =1$ and to AdS boundary conditions and therefore does not depend on any free parameter in TMG.

Let us now comment on the role of the non-Einstein solutions to the quantum theory of chiral gravity. Building on the work of Witten \cite{Witten:2007kt}, the authors of \cite{Maloney:2009ck} showed that summing only the \emph{real} saddle points of the Euclidean action amounts to consider only Einstein metrics, because the euclidean equations of motion for TMG have the simple form
\begin{equation}\label{EuclEOM}
 {\cal G}_{\mu \nu} + i \ell C_{\mu \nu} = 0.
\end{equation}
The fact these Einstein metrics are all locally isometric to three dimensional hyperbolic space drastically simplifies the computation of the sum over geometries with a two-torus as conformal boundary. The regularized sum then coincides with a chiral half of an extremal CFT partition function \cite{Witten:2007kt,Manschot:2007ha}. It was also noted that all non-Einstein solutions of TMG admit a complex analytical continuation as can be seen from \eqref{EuclEOM}. Now, the existence of Lorentzian non-Einstein solutions of TMG opens the possibility that complex geometries might play a role in the euclidean path integral, as would do e.g. the complex Euclidean Kerr geometry in four-dimensional gravity. The question of regularity is crucial since the path integral is usually restricted to \emph{smooth} geometries: for instance, negative mass black holes -- corresponding to conical defects -- are not included. Moreover, when one considers a boundary torus as conformal completion, the geometry should be well-defined everywhere which constraints the potential complex saddle points, e.g. the analytic continuation of the solution \eqref{eq:t} cannot be defined on the torus since it would be multi-valued.

The existence of regular complex saddle points with a conformal boundary torus is speculative but plausible since any anzatz $F(t,\phi) = e^{n t}e^{i m \phi}$ in \eqref{asymptsol} would lead to such a well-defined saddle point upon analytic continuation, at least close to the boundary. The classical regularized Euclidean action $I_{cl}$ of such saddle points can be computed using the arguments of \cite{Kraus:2006nb}. Since the boundary stress-tensor does not depend on the function $F(t,\phi)$ and the diffeomorphism anomaly does not play a role on the boundary torus, we obtain $e^{-I_{cl}} = \bar q^{\bar L_0}$ for any solution of the form \eqref{asymptsol}. The boundary gravitons can be included as well with the standard arguments. The partition function would then factorize 
\be
Z(\tau)  = Tr\left( \bar q^{\bar L_0} \right) = \mathcal N \,\bar Z(\tau)
\ee
where $\bar Z(\tau)$ is the anti-holomorphic extremal partition function described in \cite{Maloney:2009ck} and $\mathcal N$ is a (regularized) degeneracy due to the complex saddle points\footnote{If it turns out that the only regular complex saddle points have $\bar L_0 = 0$, the inclusion of the latter would reflect only in the change of the constant term in the extremal partition function which usually only counts descendants of the vacuum, see e.g. (3.8) of \cite{Witten:2007kt}.}. This leads us to conclude that even if such saddle points exist, the partition function of chiral gravity is still likely to be chiral and modular invariant.

\subsection*{Acknowledgements}
We thank Dionysios Anninos, Gaston Giribet, Daniel Grumiller, Monica Guica, Gary Horowitz, Niklas Johansson, Alex Maloney, Don Marolf, Ivo Sachs, Kostas Skenderis, Wei Song and Andy Strominger for many useful discussions and correspondences. This research was  supported in part by the National Science Foundation under Grant No. PHY05-51164. The work of G.C. is supported in part by the US National Science Foundation under Grant No.~PHY05-55669, and by funds from the University of California. The works of S.dB. and S.D. are funded by the European Commission though the grants PIOF-GA-2008-220338 and PIOF-GA-2008-219950  (Home institution: Universit\'e Libre de Bruxelles, Service de Physique Th\'eorique et Math\'ematique, Campus de la Plaine, B-1050 Bruxelles, Belgium).


\providecommand{\href}[2]{#2}\begingroup\raggedright\endgroup

\end{document}